\documentclass[a4paper]{article}

\usepackage{INTERSPEECH2020}
\usepackage{color,soul}
\usepackage{adjustbox}
\usepackage{array,multirow,graphicx}
\usepackage{float}
\usepackage{hyperref}
\usepackage[table]{xcolor}

\title{Controllable Neural Prosody Synthesis}
\name{Max Morrison$^{1,*}$\thanks{$^*$This work was carried out during an internship at Adobe Research.}, Zeyu Jin$^{2}$, Justin Salamon$^{2}$, Nicholas J. Bryan$^{2}$, Gautham J. Mysore$^{2}$}
\address{$^{1}$Northwestern University, Evanston, IL, USA\\
$^{2}$Adobe Research, San Francisco, CA, USA}
\email{maxrmorrison@gmail.com, zejin@adobe.com}

\newcolumntype{R}[2]{%
    >{\adjustbox{angle=#1,lap=\width-(#2)}\bgroup}%
    l%
    <{\egroup}%
}

\begin{document}

\maketitle
\begin{abstract}
Speech synthesis has recently seen significant improvements in fidelity, driven by the advent of neural vocoders and neural prosody generators. 
However, these systems lack intuitive user controls over prosody, making them unable to rectify prosody errors (e.g., misplaced emphases and contextually inappropriate emotions) or generate prosodies with diverse speaker excitement levels and emotions.
We address these limitations with a user-controllable, context-aware neural prosody generator. Given a real or synthesized speech recording, our model allows a user to input prosody constraints for certain time frames and generates the remaining time frames from input text and contextual prosody. We also propose a pitch-shifting neural vocoder to modify input speech to match the synthesized prosody. Through objective and subjective evaluations we show that we can successfully incorporate user control into our prosody generation model without sacrificing the overall naturalness of the synthesized speech.

\end{abstract}

\section{Introduction}
\label{sec:intro}
Text is a strong indicator of prosodic patterns \cite{stanton2018predicting}, but not a determinant. For the same text, prosody varies with speaker intention \cite{wells2006english}, which imposes challenges for modern text-to-speech models \cite{shen2018natural, ping2017deep, NIPS2019_8580} in the form of misplaced emphases and degraded naturalness. Manually attempting such corrections using an audio editor, as is done in podcast and video dialogue editing, requires expertise in speech manipulation and significant time and effort. In this paper, we aim to address these issues by proposing an intuitive and less error-prone process, consisting of three steps: (1) a user provides constraints on the prosody (e.g., by drawing part of an F0 contour), (2) a neural prosody generator predicts an F0 contour for the whole utterance while matching the user's constraints, and (3) a neural vocoder synthesizes a high-fidelity speech recording that exhibits the generated prosody.

Early approaches for F0 synthesis use techniques such as decision trees \cite{dusterhoff1999using}, unit selection \cite{raux2003unit}, and hidden Markov models (HMMs) \cite{yu2010continuous}. More recently, deep learning methods such as variational autoencoders (VAEs) \cite{hodari2019using}, deep autoregressive (DAR) neural networks \cite{wang2018autoregressive}, and vector-quantized VAEs (VQ-VAEs) \cite{wang2019vector} were shown to be effective at generating F0 contours of speech from text features. Hodari et. al. \cite{hodari2019using} show that VAEs produce F0 contours that cluster around placing emphasis on the same words despite repeated sampling. This indicates that the VAE is not capturing the multimodal nature of English prosody associated with contrastive emphases. For example, the sentence ``the dog is black'' communicates a different intention when one of ``dog'', ``is'', or ``black'' is emphasized. The DAR model proposed in \cite{wang2017rnn} has previously shown promise in modeling the multimodality of English prosody \cite{wang2018autoregressive} but does not allow user control over F0 generation. While our work focuses on user control of F0 generation, additional prosodic control can be achieved by first generating a speech waveform with the desired phoneme durations (e.g., with \cite{NIPS2019_8580}) and then using our method to achieve the desired F0. 

Once we generate an F0 contour, we synthesize speech using a compatible vocoder. Existing vocoders allow either perceptually high-quality synthesis \cite{shen2018natural, ping2019waveflow, wang2019neural} or a high degree of control over prosody \cite{morise2016world}. Recently, significant effort has gone into disentangling the latent spaces of high-quality neural vocoders to recover explicit prosodic control \cite{hsu2018hierarchical, lee2019robust, habib2019semisupervised, sun2020fullyhierarchical}. One such model, Quasi-Periodic WaveNet, allows frame-wise F0 control via an explicit F0 contour but produces lower naturalness than DSP-based vocoders, especially when pitch is shifted upward \cite{Wu2019}. In contrast, we propose a pitch-shifting neural vocoder that achieves comparable or superior performance as DSP-based methods while factorizing prosody control parameters in the input space using a simple, jointly-trained bottleneck.

Our key contributions are: (1) a novel method for F0 generation that permits intuitive user controls, (2) a pitch-shifting neural vocoder with explicit F0 conditioning, and (3) a new subjective evaluation method for measuring the naturalness of prosody. Through our perceptual evaluation, we show that user control of prosody can be obtained without degrading prosody naturalness, and our pitch-shifting neural vocoder performs comparably with existing DSP-based methods while outperforming prior neural pitch-shifting methods.

\section{Controllable F0 generation}
\label{sec:metho}
\subsection{Deep autoregressive (DAR) neural network}
For its effectiveness and simplicity, we use DAR as our baseline model for F0 generation (model $\texttt{QF}_{\texttt{FT}}$ \cite{wang2017rnn}). DAR feeds input text features through two fully-connected layers with ReLU activation followed by two RNNs, one bidirectional and one unidirectional, followed by a fully-connected layer. The outputs of the last fully-connected layer are the logits of a categorical distribution of quantized F0 values. One F0 value is sampled per time frame and the resulting observation (a one-hot-encoded F0 value) is concatenated to the input of the unidirectional RNN at the next frame (i.e., the unidirectional RNN is autoregressive). To prevent the unidirectional RNN from ignoring the current input features and focusing on its hidden state (i.e., exposure bias), \textit{data dropout} is used, whereby autoregressive inputs to the unidirectional RNN are set to zero with probability $p$ (we use $p=0.5$, as in \cite{wang2017rnn}). The original DAR uses a hierarchical softmax loss to improve binary classification of voiced/unvoiced (V/UV) frames. However, the ground truth V/UV sequence can also be derived directly from phonemes when synthesizing speech, or from a preexisting F0 contour when editing speech. We use V/UV sequences from existing F0 contours and therefore use cross-entropy loss instead of hierarchical softmax.

\begin{figure}[t]
    \centering
    \includegraphics[width=\linewidth]{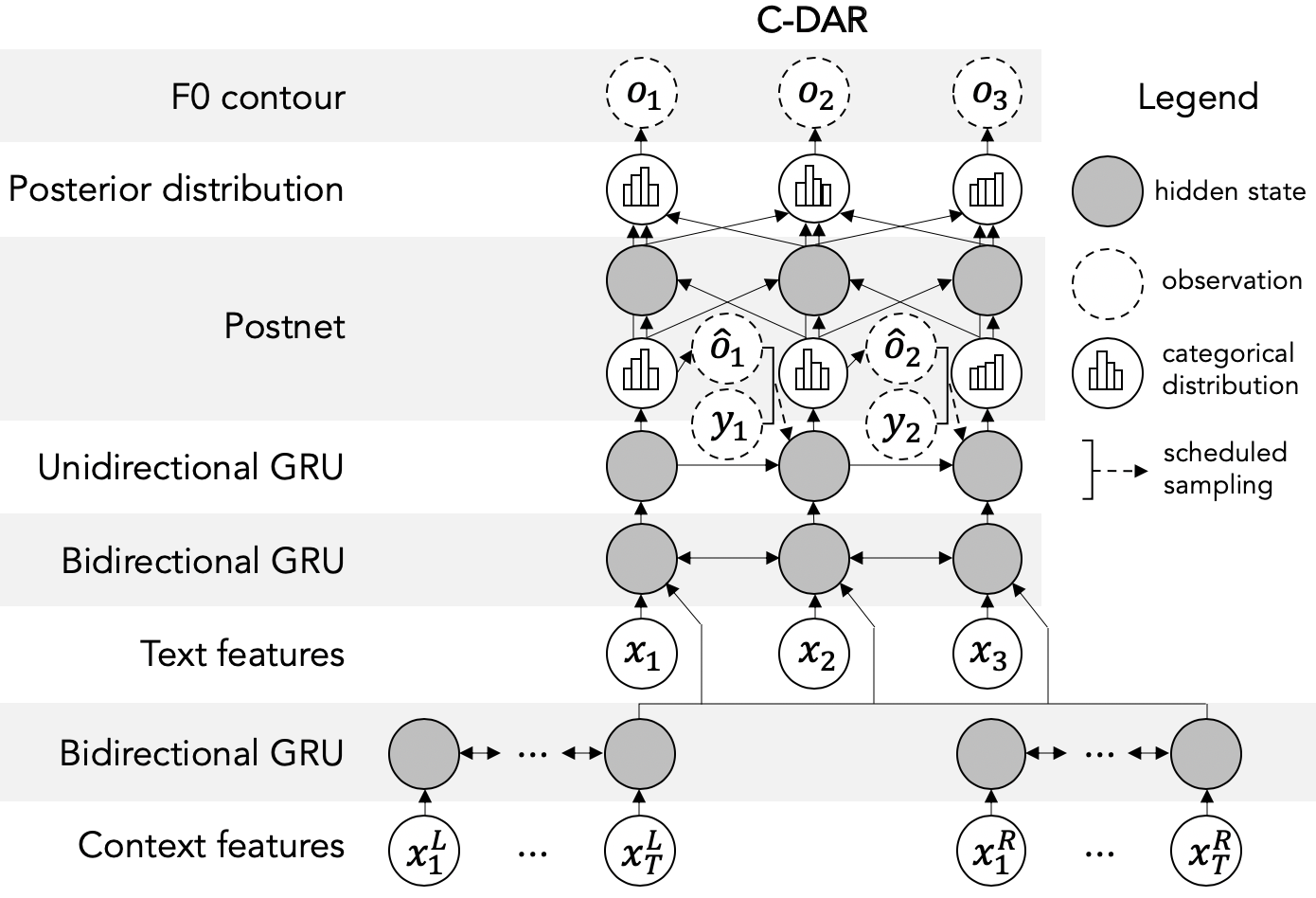}
    \caption{Our proposed C-DAR architecture for controllable F0 generation. $x_t, y_t, o_t, $ and $\hat{o}_t$ are the input features, ground truth F0, predicted F0, and predicted F0 before postnet, respectively, and $t = 1, \dots , T$ is the current frame. $x_t^L$ and $x_t^R$ are the input features from the preceding and following speech. The fully-connected layers between the text features and the bidirectional GRU as well as three layers of the postnet are omitted.}
    \label{fig:models}
\end{figure}

DAR has been shown to be effective at modeling English prosody, but does not permit user control and lacks the context-awareness necessary for speech editing tasks. We address these limitations in our proposed F0 generation model, Controllable DAR (C-DAR), shown in Figure \ref{fig:models}.

\subsection{Controllable DAR}
A significant advantage of working with an F0 contour, as opposed to jointly predicting all prosodic features, is that users may explicitly create, modify, and constrain the F0 contour to realize a creative goal. We propose three techniques designed to facilitate control of a DAR-based model for F0 generation: (1) if available, the preceding and following speech content is summarized and used to condition the model, (2) random segments of the ground truth F0 are provided to the model during training, and (3) the model predicts F0 values in reverse order.

The preceding and following speech content provides useful indicators for placing emphases \cite{wells2006english}, capturing the speaker's current prosodic style \cite{im2018probabilistic}, and determining F0 values near boundary points. This context-awareness is essential in speech editing tasks, where prosody edits must sound natural relative to the surrounding speech. We incorporate context-awareness by summarizing the preceding and following content each with an untied, two-layer bidirectional GRU with hidden size 128. We use the same input features for the preceding and following content (see Section \ref{sec:feats}) with the addition of one-hot-encoded F0 values. The result is concatenated with the text features at the input of the model at each time frame.

\begin{figure}[t]
    \centering
      \centerline{\includegraphics[width=\linewidth]{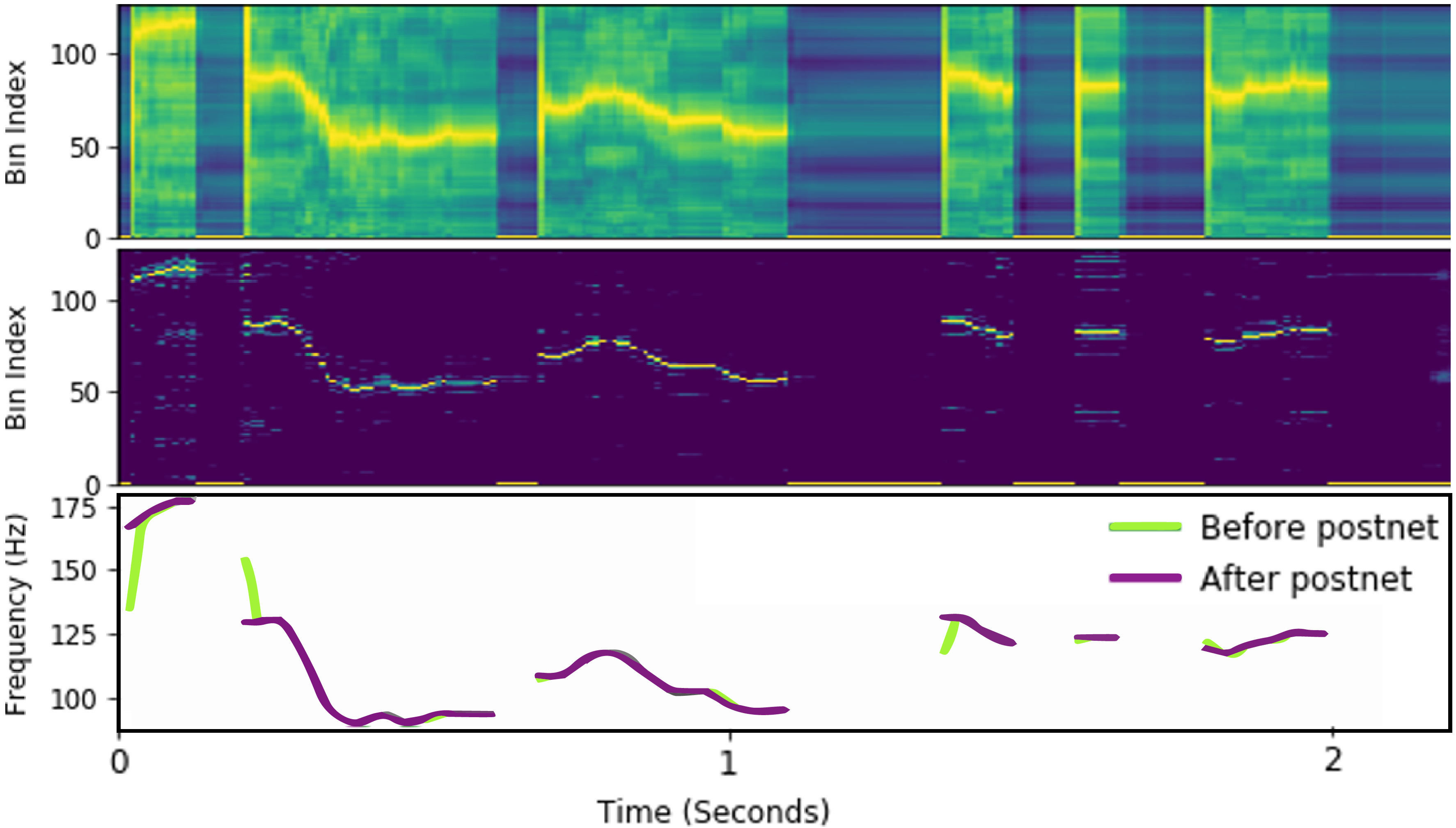}}
    \caption{Effects of postnet on F0 generation. \textbf{(Top)} Log posterior distribution at each frame before postnet. \textbf{(Middle)} Log posterior distribution after postnet. \textbf{(Bottom)} Argmax of each distribution converted to frequency. Posteriors are per-frame normalized to have a maximum of 0 and clipped below -40.}
    \label{fig:postnet}
\end{figure}

A potentially useful user interaction for controllable F0 generation is to explicitly specify some segments of the desired contour (e.g., by placing and moving anchor points or drawing) and have a generative model infer the remaining F0 values. This permits iterative refinement, in which a user generates an F0 contour using our model, selects regions they want to keep, regenerates only unselected regions, and repeats until satisfied. Explicitly specifying a higher pitch over a word also allows users to quickly create emphases\footnote{We demonstrate this use case and provide audio examples of our experiments at \url{https://www.maxrmorrison.com/sites/controllable-prosody}.}. This is useful as emphases in English are challenging to predict \cite{mass2018word} as they can arise semantically \cite{gussenhoven2008types} or simply due to speaker excitement \cite{im2018probabilistic}. We implement this technique by providing ground-truth F0 as an input feature during training for random subsequences between 10 and 1000 milliseconds, and explicitly conditioning the autoregressive RNN on this ground truth F0. Explicitly conditioning the unidirectional RNN allows it to predict the correct F0 with high accuracy while, we hypothesize, the input features encourage the model to learn to generate smooth, continuous F0 surrounding the specified contour rather than suddenly jumping to the specified contour. During inference, we use the user-specified F0 values instead of the ground-truth F0. Because a recurrent model uses a combination of its input features and history and does not have a reliable history at the start of generation, discontinuities are reduced when the specified contour occurs towards the start of generation. Therefore, downstream tasks that require more editing in the second half of an utterance benefit from reversing the order of sequence generation.

Relative to DAR, C-DAR has three additional changes that do not significantly impact naturalness or controllability, but provide additional insights into F0 generation. First, a 5-layer postnet \cite{shen2018natural} follows the autoregressive RNN. We find that this postnet has the effect of reducing autoregressive sampling errors and tightening the posterior distribution around the argmax (Figure \ref{fig:postnet}). Second, we use scheduled sampling \cite{bengio2015scheduled} instead of data dropout. Scheduled sampling is known to not be a consistent estimator \cite{huszar2015not} and was shown to exhibit worse objective metrics for F0 generation (V/UV precision and pitch RMSE) \cite{wang2017rnn}. Our findings indicate that neither consistency nor superior objective metrics are reliable indicators of improved subjective naturalness. Lastly, our bidirectional RNN has 16 hidden units instead of 256, indicating that prior F0 generation models may be using more model capacity than necessary.

\subsection{Input features for F0 generation}
\label{sec:feats}
For each 10 ms frame, we concatenate five input features: (1) the one-hot-encoded phoneme, (2) a BERT \cite{devlin2018bert} word embedding, (3) the V/UV label, and (4) one-hot encodings of nearby punctuation (e.g., whether the word precedes a comma or is in quotations). We use P2FA \cite{yuan2008speaker} for phoneme alignment. Word embeddings are computed by averaging over subword tokens extracted via the \texttt{bert-large-uncased} pretrained model from the HuggingFace Transformers package \cite{Wolf2019HuggingFacesTS}. These features are jointly referred to as ``Text features'' in Figure 1.

\subsection{F0 representation}
\label{sec:pitch}
We use a modified version of CREPE \cite{kim2018crepe} to extract ground truth F0 contours. Our modification is as follows: rather than performing a localized search around the argmax of the posterior distribution over F0 bins, we directly decode the F0 contour from the posterior distribution via Viterbi decoding. Our transition matrix places maximal probability on maintaining the same F0 value and zero probability on F0 discontinuities greater than 240 cents---with linearly decreasing probability in between. We determine V/UV labels via hysteresis thresholding applied to CREPE's harmonicity confidence value. During training, we quantize our F0 representation to one of 128 values. We reserve one value for predicting unvoiced tokens and evenly divide the other 127 bins to span $4$ standard deviations above and below the speaker's average F0 in base-2 log-space.



\begin{table}
\centering
\begin{tabular}{rllc}
\textbf{ID} & \textbf{Gender} & \textbf{Book(s)} & \textbf{Hours} \\
\hline
\hline
\textbf{94}   & Male   & \textit{The Canterville Ghost} & 3.00 \\
\textbf{3906} & Female & \textit{Little Fuzzy}          & 4.19 \\
\textbf{5717} & Male & \textit{The Cowardly Lion of Oz} & 4.19 \\
\textbf{11049} & Female & \textit{The Warren Report} & 7.76 \\
\end{tabular}
\caption{Speakers used for evaluation and vocoder training.}
\label{tab:speak}
\end{table}

\section{Pitch-shifting WaveNet vocoder}
In order to synthesize speech from an arbitrary F0 contour, we propose a pitch-shifting WaveNet vocoder (PS-WaveNet) that accepts an F0 contour as conditioning. We use a publicly available implementation \cite{r9y9wavenet} of a single-speaker WaveNet vocoder \cite{tamamori2017speaker} and predict 10-bit $\mu$-law-encoded waveform samples. We enable explicit F0 control by forcing the input acoustic features through a small, jointly learned bottleneck that encourages the network to take F0 features solely from the input F0 contour.

We use 21-channel mel-cepstral coefficients (MCeps) instead of the typical 80-channel log-mel-spectrograms as MCep are less individually representative of energies at specific frequencies and therefore more easily separable from F0. Our informal experiments using 80-channel log-mel-spectrograms produced samples that were relatively inharmonic, but also captured more high-frequency detail. Our MCep bottleneck consists of three 1D convolutional layers with a filter width of 5 and ReLU activations between layers. The output channels of each convolutional layer are 20, 20, and 12, respectively. We train our PS-WaveNet vocoder on the bottlenecked MCep features concatenated with the one-hot-encoded F0 contour.

\section{Experimental Design}

We use the 360-hour clean training data partition from LibriTTS dataset \cite{zen2019libritts} to train our F0 generation networks. We train our PS-Wavenet vocoders and evaluate all models using LJSpeech \cite{ljspeech17} as well as three single-speaker datasets similarly constructed from LibriVox recordings. The reader ID, gender, book, and amount of training data for each speaker are given in Table \ref{tab:speak}. For evaluation, we use 20 held-out utterances (10 questions and 10 statements) from each speaker with duration between 2 and 10 seconds. For all subjective listening tasks, we collect 25 responses from each of our 48 US participants on Amazon Mechanical Turk (AMT).

\subsection{Model training}
We train DAR and C-DAR with a batch size of 32 utterances and an ADAM optimizer \cite{kingma2014adam} with a learning rate of $10^{-3}$. We find that validation loss corresponds poorly with naturalness. Instead of early stopping, we train for 15 epochs (3.6k steps per epoch) and manually listen to results from the LibriTTS validation set from epochs 5-15. We find that DAR and C-DAR produce the most natural F0 contours after 9 epochs. We train one PS-WaveNet for each speaker in Table \ref{tab:speak}. We train for 475k steps with a batch size of 8 and an ADAM optimizer with a learning rate of $10^{-3}$ that is halved every 200k steps. We use 30 dilated convolution layers with dilation rate $2^{\ell \texttt{ mod } 10}$ at layer $\ell$. Noise injection with $10^{-3}$ Gaussian noise is used to improve the stability of synthesis \cite{Jin:2018:FAR}.


\subsection{Evaluating F0 contour generation}
\label{sec:evdar}
We evaluate F0 generation models using both objective and subjective metrics, but emphasize that subjective metrics align best with our goal. Our objective metrics are the pitch RMSE and the negative log-likelihood (NLL) of the model relative to ground truth pitch. We do not report V/UV metrics, as all models correctly preserve the V/UV sequence. We report two subjective metrics, including a novel subjective metric that addresses a problem with previous F0 generation evaluation methods.

Prior work on F0 generation uses pitch-shifting vocoders to generate evaluation samples, which participants listen to and provide a naturalness rating \cite{hodari2019using, wang2018autoregressive, wang2019vector}. Here, we address the issue where artifacts induced by pitch-shifting are proportional to the size of the shift. This penalizes natural-sounding F0 contours that have high $\ell_1$ or $\ell_2$ distance from the original pitch, and rewards unnatural F0 contours close to the original pitch. To address this, we low-pass filter each vocoded sample at 10 Hz above the maximum F0. The resulting audio preserves the F0 contour and amplitude envelope while removing all artifacts above the cutoff frequency. During the user study, participants are told that they are listening to the intonation of speech spoken by either a real person (``real'') or synthesized by a computer (``fake''), and are asked to identify each sample as real or fake.

We implement our proposed user study to evaluate the naturalness of DAR and C-DAR. We use the PSOLA vocoder, and include as baselines a monotone model as well as two random models: \textit{replace}, which replaces the F0 contour of each word with a contour from a random word uttered by the same speaker, and \textit{swap}, which randomly swaps F0 contours of words within the sentence. For completeness, we also conduct the more typical MOS naturalness test without low-pass filtering using our proposed vocoder (see Section \ref{sec:evwav}).

Our second subjective study evaluates the controllability of the C-DAR model on the task of synthesizing F0 after changing a question mark to a statement, or vice versa. We call this task ``repunctuation''. Our weak baseline is the original audio with the original punctuation. As a strong but unnatural baseline, we replace only the last two words of a sentence with a manually-selected F0 contour that is representative of the target punctuation. For DAR and C-DAR, we change the punctuation of the text input. For C-DAR, we also provide the F0 of the last two words as a user-specified F0 segment. Samples are vocoded using PSOLA and low-pass filtered as described above. AMT participants are given a sample and asked to select whether the sample sounds more like a statement or question.

\subsection{Evaluating PS-WaveNet}
\label{sec:evwav}
We evaluate the consistency and naturalness of PS-WaveNet via two tasks. For both tasks, our baselines are PSOLA \cite{charpentier1986diphone} and WORLD \cite{morise2016world}---two DSP-based vocoders with frame-wise F0 control. For our first task, we measure how closely the synthesized speech follows the given F0 contour via the F0 RMSE and V/UV errors between the input and output F0. We obtain the F0 of the output using our method described in Section \ref{sec:pitch}. For our second task, AMT participants rate the naturalness of each sample between 1 (low naturalness) and 5 (high naturalness). We evaluate all vocoders using both the original F0 contour and the F0 contour generated by C-DAR. We include the original audio and intentionally degraded audio (quantized to 3 bits) as references for high and low naturalness, respectively.

\begin{table}
\centering
\begin{tabular}{lccc}
\textbf{F0 source}       & \textbf{NLL} & \textbf{RMSE} & \textbf{\% Considered Real} \\
\hline
\hline
Original           & --  & 0.00 & 0.72 \\
\rowcolor{gray!12}
Monotone               & --  & 0.37 & 0.19 \\
Random (swap)      & --  & 0.37 & 0.37 \\
\rowcolor{gray!12}
Random (replace)   & --  & 0.43 & 0.38 \\
DAR                & 8.15 & 0.43 & 0.57 \\
\rowcolor{gray!12}
C-DAR            & 9.97 & 0.45 & 0.55
\end{tabular}
\caption{Results for objective F0 generation experiments and the subjective low-pass experiment. Lower scores are better for NLL and RMSE and higher is better for \% Considered Real.}
\label{tab:hum}
\end{table}

\begin{table}
\centering
\begin{tabular}{llccc}
\textbf{F0 source} & \textbf{Vocoder} & \textbf{V/UV Metrics} & \textbf{RMSE} & \textbf{MOS} \\
\hline
\hline
3-bit & None  & 0.99/0.64 & 0.09 & 1.48 \\
\rowcolor{gray!12}
Original & None & 1.00/1.00 & 0.00 & 4.30 \\
Original         & PSOLA & 0.99/0.98 & 0.06 & 4.10 \\
\rowcolor{gray!12}
Original         & WORLD & 0.97/0.80 & 0.05 & 3.61 \\
Original         & PS-WN & 0.93/0.87 & 0.25 & 3.73 \\
\rowcolor{gray!12}
C-DAR            & PSOLA & 0.98/0.97 & 0.19 & 3.55 \\
C-DAR            & WORLD & 0.97/0.80 & 0.07 & 3.11 \\
\rowcolor{gray!12}
C-DAR            & PS-WN & 0.93/0.85 & 0.32 & 3.52 \\
DAR              & PS-WN & 0.94/0.84 & 0.28 & 3.41 \\
\end{tabular}
\caption{Pitch-shifting vocoder experiment results. PS-WN is our proposed PS-WaveNet. V/UV metrics are formatted as \texttt{precision/recall}.}
\label{tab:vocod}
\end{table}
\section{Results}

\subsection{F0 Generation}
We present the F0 generation results in Table \ref{tab:hum}. We see that C-DAR achieves a comparable naturalness to DAR while enabling user control and context-awareness. This is further corroborated by the mean opinion scores (MOS) in Table \ref{tab:vocod}, which show that participants considered C-DAR to be slightly more natural than DAR. The results of Table \ref{tab:hum} also corroborate that NLL and RMSE are unsuitable metrics for F0 generation: neither correlates with subjective perceptions of naturalness. Further, we found that NLL could be trivially lowered by training C-DAR for fewer epochs, but with clearly degraded naturalness. This reinforces the need for domain-specific subjective metrics such as our proposed low-pass evaluation method.

We present our repunctuation experiment results in Table \ref{tab:repun}. Relative to DAR, using C-DAR with short, user-specified F0 contours improves the adherence of the generated F0 contour to high-level semantic concepts (e.g., questions and statements). We find this to be especially true when the target punctuation is a question mark. We believe this is because statements are heavily over-represented in the dataset, leading to class-imbalance and mode collapse. Our results indicate that simple user inputs make for an effective mode selector for prosody generation.

\begin{table}
\centering
\begin{tabular}{ccccc}
 & \textbf{Original} & \textbf{Heuristic} & \textbf{Monotone} & \textbf{DAR} \\
\textbf{Original} & - & - & - & - \\
\rowcolor{gray!12}
\textbf{Heuristic} & 0.55/0.54 & - & - & - \\
\textbf{Monotone} & - & 0.28/0.36 & - & - \\
\rowcolor{gray!12}
\textbf{DAR} & 0.43/0.60 & 0.41/0.46 & 0.68/0.49 & - \\
\textbf{C-DAR} & 0.59/0.60 & 0.49/0.46 & 0.71/0.69 & 0.63/0.50 \\
\end{tabular}
\caption{Repunctuation experiment results. A pairwise comparison of five models. All results indicate percent preference for the model specified in the same row over the model in the same column. Results are formatted as \texttt{Q/S}, where \texttt{Q} and \texttt{S} are the percent preferences when the target punctuations are question marks and periods, respectively. \textbf{Heuristic} is our strong baseline described in Section \ref{sec:evdar}.}
\label{tab:repun}
\end{table}

\subsection{PS-WaveNet}
In Table \ref{tab:vocod}, we see that our PS-WaveNet significantly outperforms the naturalness of WORLD while achieving comparable performance to PSOLA. We find that PS-WaveNet has a higher variance of MOS across speakers, ranging from 3.04 for speaker 5717 to 3.80 for speaker 94 when using C-DAR. In comparison, PSOLA achieves 3.40 and 3.58 MOS on speakers 5717 and 94, respectively. An additional pairwise test between PS-WaveNet and PSOLA using F0 contours generated with C-DAR confirms that PSOLA is preferred only for speakers 5717 and 11049.

The objective metrics reported in Table \ref{tab:vocod} highlight additional tradeoffs when selecting a pitch-shifting method. For example, we see that WORLD achieves the best RMSE despite its low MOS, but also tends to make unvoiced regions sound voiced (i.e., low V/UV recall). This is more useful for pitch-shifting singing, for example, as high pitch accuracy is important and unvoiced regions are less common than speech. PS-WaveNet achieves higher V/UV recall than WORLD, but at a cost to V/UV precision and RMSE. We hypothesize that the increase in inharmonicity due to lower V/UV precision also induces more pitch-tracking errors, including pitch-doubling errors which produce extremely high RMSE.

\section{Conclusion}
In this work, we present a deep autoregressive model that supports controllable, context-aware F0 generation; a pitch-shifting neural vocoder that allows explicit F0 conditioning; and novel subjective evaluation methods for F0 generation. We show in user studies that our controllable F0 model exhibits comparable naturalness as non-controllable baselines, and that our pitch-shifting neural vocoder exhibits comparable naturalness as DSP-based vocoders. There are many directions for future work, including real-time pitch-shifting vocoding and interaction design for prosody editing.

\bibliographystyle{IEEEtran}

\bibliography{bibliography}

\end{document}